\documentstyle[prd,aps]{revtex}

\begin{document}
\preprint{LANCS-TH/9505, RESCEU-16/95, hep-ph/9510204}
\draft
\tighten

\title{Thermal Inflation and the Moduli Problem}
\author{David H. Lyth}
\address{School of Physics and Chemistry, Lancaster University,
Lancaster, LA1 4YB, U.K.}
\author{Ewan D. Stewart}
\address{Research Center for the Early Universe, School of Science,
University of Tokyo, Bunkyo-ku, Tokyo 113, Japan}
\date{September 1995}
\maketitle
\begin{abstract}
In supersymmetric theories a field can develop a vacuum expectation value
$M \gg 10^3\,{\rm GeV}$, even though its mass $m$ is of order $10^2$ to
$10^3\,{\rm GeV}$.
The finite temperature in the early Universe can hold such a field at zero,
corresponding to a false vacuum with energy density $ V_0 \sim m^2 M^2 $.
When the temperature falls below $V_0^{1/4}$, the thermal energy density
becomes negligible and an era of thermal inflation begins.
It ends when the field rolls away from zero at a temperature of order $m$,
corresponding to of order 10 $e$-folds of inflation which does not affect
the density perturbation generated during ordinary inflation.
Thermal inflation can solve the Polonyi/moduli problem if
$ M $ is within one or two orders of magnitude of $10^{12}\,{\rm GeV}$.
\end{abstract}

\section{Introduction}

There is at present a `standard model' of the Universe before nucleosynthesis,
which is described in many reviews and several textbooks. According to this
model, an early era of inflation sets the initial conditions for a Hot Big
Bang, which starts far above the critical temperature for the electroweak
transition ($T\simeq 100\,{\rm GeV}$) and continues without interruption
until the
present matter dominated era begins.

This picture is pleasingly simple, but it is by no means mandatory in the
context of current thinking about the fundamental interactions
beyond the Standard Model. To be precise, it will not be valid if
one or more scalar fields have a sufficiently
large vev (vacuum expectation value) while at
the same time having an almost flat potential. The reason is that
the particle species corresponding to the oscillation around such
a vev is typically both abundant and long lived, which modifies the
simple picture in a significant and sometimes disasterous way.
Extending an old terminology \cite{decay}, we shall call a scalar field
with a large vev and a flat potential a `flaton field', or simply a
`flaton'.\footnote
{Note the etymology. The term `flaton' refers to the {\em flat\/} potential,
not to in{\em flat\/}ion. Conversely, the familiar word `inflaton' refers to
the field which is slowly rolling during {\em inflat\/}ion.
We shall also use the term `flaton'
to denote the particle species corresponding to a flaton field.}

Although flaton fields are by no means inevitable, they are natural
in the context of modern particle theory and in our opinion their
possible cosmological consequences should be taken very seriously.
Some aspects of the cosmology of flaton fields are already well known
\cite{decay,dinefisch,coughlan,yam,therm,Enqvist,Ross,yam2,interm},
and in a recent note \cite{gutti} we drew attention to a new feature
which we termed `thermal inflation'.
The present paper, along with two more in preparation \cite{david,ewan},
aims to give a systematic account of the subject.

Let us begin by being more precise about what is meant
by a `large' vev, and a potential which is `almost flat'.
These terms are defined with respect to the energy scale $10^2$ to
$10^3\,{\rm GeV}$, which is the scale of supersymmetry breaking as defined by
the masses of the supersymmetric partners of known particles \cite{susy}.
The vev is defined as the position of the minimum of the potential,
and a `large' vev $M$ is one satisfying $M\gg 10^3\,{\rm GeV}$.
An `almost flat' potential $V$ is one whose curvature $|V''|^{1/2}$
is of order $10^2$ to $10^3\,{\rm GeV}$ (except near any points of inflexion)
out to field values much bigger than $10^3\,{\rm GeV}$,
and if the field has a large vev this is supposed to be true
out to at least the vev.
For an almost flat potential the particle mass $m$
is therefore of order $10^2$ to $10^3\,{\rm GeV}$.
{}From now on we drop the qualifier `almost', referring simply
to a flat potential.

The most widely discussed flaton candidates are the moduli occurring in
superstring theory. The potential of a modulus is indeed flat, and if
its vev is nonzero it is typically of order the Planck scale
$M_{\rm Pl}=(8\pi G)^{-1/2}=2.4\times 10^{18}\,{\rm GeV}$.
A modulus with such a vev\footnote
{A field with these properties occurred in the first
example \cite{Polonyi} of a nonrenormalizable supersymmetry-breaking hidden
sector, which contained a single complex field. It was called the Polonyi
field, and the associated problem \cite{problem} was called the Polonyi
problem. Most of what we say concerning the moduli applies to any species
with these properties.}
is known to be fatal to the standard cosmology since the
corresponding particles are very abundant and do not decay before
nucleosynthesis \cite{problem,dilaton,Banks,Randall,Steinhardt,Dine}.
As we shall see, the failure to decay before nucleosynthesis is likely
to persist for any flaton with a vev exceeding $10^{14}\,{\rm GeV}$,
making all such flatons fatal to the standard cosmology \cite{Enqvist}.

Moduli are by no means the only flaton candidates. On the contrary,
{\em any\/} field (in the observable sector)
with a vev much bigger than $10^3\,{\rm GeV}$
is likely to have a flat potential, and so to be a flaton.
The reason, as we discuss in detail below, is that it is natural to
construct all available mass scales from just the two basic scales
$m$ and $M_{\rm Pl}$. Apart from the moduli, the
most familiar examples of fields with nonzero vevs
are those which are charged under a continuous symmetry,
the vev then indicating a spontaneous breakdown of the symmetry.
If the symmetry is local then the field is by definition a higgs
field, and presumably the examples of this type occurring in nature
(apart from the higgs fields breaking electroweak symmetry)
are the higgs fields breaking the GUT symmetry, whose
vevs are of order $10^{16}\,{\rm GeV}$.
Alternatively the symmetry could be global,
a likely candidate for this case being the Peccei-Quinn field
with a vev perhaps of order $10^{11}\,{\rm GeV}$.
On the other hand, it makes perfect sense for a field to have a
nonzero vev even if it is not charged under any continuous symmetry.
For example, a right-handed neutrino mass might be generated by
a vev, without lepton number being a good symmetry \cite{flataxion,ewan}.

As mentioned already, moduli as well as any
other flatons with a vev bigger than
$10^{14}\,{\rm GeV}$ are fatal
to the standard cosmology.
How are we to solve this `moduli problem' if it exists?

The usual recipe for getting rid of unwanted relics in cosmology
is to invoke
an early epoch of inflation, lasting at least 50 to 60 Hubble times or
so. Such an era is also desirable for other reasons \cite{Linde,KT},
one of which is that it
can generate an adiabatic density perturbation of the right
magnitude to explain the cosmic microwave background anisotropy and large
scale structure.
To do this the potential at the end of inflation must
satisfy $ V^{1/4} \lesssim 10^{16}\,{\rm GeV} $ \cite{bound},
and the lowest value of $ V^{1/4} $ that has been proposed in a
plausible model is $ V^{1/4} \sim 10^{12}\,{\rm GeV} $ \cite{hybrid,fvi}.

Inflation at such a high scale does not solve the moduli problem, because
although it sufficiently dilutes moduli present before inflation
they are regenerated with an unacceptable abundance afterwards.
We show in \cite{gutti}, and in much more detail below,
that to avoid excessive regeneration one requires
\begin{equation}
V^{\frac{1}{4}} \lesssim 10^7 \,\mbox{to}\, 10^8\,{\rm GeV}
\left( \frac{\mbox{GeV}}{T_{\rm R}} \right)^{\frac{1}{4}}
\label{vbound}
\end{equation}
where $T_{\rm R}$ is the reheat temperature.
An era of inflation at such a low energy scale seems impossible
to realize in the context of sensible particle physics, if it is
required also to produce the cosmological density perturbation.
Randall and Thomas \cite{Randall} therefore suggested that the
density perturbation is produced by an era of inflation at the usual high
energy scale, while a second era of inflation at a low energy scale
solves the moduli problem. However, even without the constraint
of producing the density perturbation it is difficult to
construct a model of inflation giving a sufficiently low energy scale,
within the usual paradigm where there is an inflaton field rolling
slowly down the potential. The reason stems from the fact
that a necessary condition for
slow roll is that the inflaton mass (or more precisely
the curvature $|V''|^{1/2}$ evaluated while the field is rolling)
be much less than the Hubble parameter $H\simeq V^{1/2}/M_{\rm Pl}$.
The bound displayed in Eq.~(\ref{vbound}) corresponds to a very low
mass $\lesssim 10\,{\rm MeV}$.

The central purpose of this paper is to explore the fact that a
flaton field can lead to a completely different type
of inflation, called thermal inflation \cite{gutti},
which can solve the moduli problem provided that the
vev $M$ is within one or two orders of magnitude of $10^{12}\,{\rm GeV}$.
During thermal inflation the flaton field is held at the origin by
finite temperature effects so that no field is rolling.
The potential during thermal inflation is the value $V_0$ of the
flaton potential at the origin, which is of order $m^2 M^2$.
With $M\sim 10^{12}\,{\rm GeV}$ this gives $V_0^{1/4}\sim 10^7\,{\rm GeV}$
which can satisfy Eq.~(\ref{vbound}).
Thermal inflation starts when the thermal energy density falls below
$V_0$ which corresponds to a temperature roughly $V_0^{1/4}$, and
it ends when the finite temperature becomes ineffective at a
temperature of order $m$, so the number of
$e$-folds is $\frac12\ln(M/m)\sim 10$. It turns out that this
can sufficiently dilute the moduli existing before thermal inflation
(especially if reheating after thermal inflation is delayed)
and it will not interfere with the density
perturbation generated during ordinary inflation. There is also the intriguing
possibility that two or more bouts of thermal inflation can occur in
quick succession, allowing an even more efficient
solution of the moduli problem.

The present paper and its two successors are complementary to recent
papers by Dine, Randall and Thomas \cite{Dine,DRTbaryo}.
The latter focus on fields with a flat potential but zero vev.
These fields too are liable to be oscillating in the early
Universe and if they carry nonzero lepton or baryon number
they can lead to baryogenesis (the Affleck-Dine mechanism).
However baryogenesis in this way works only
if there is no thermal inflation, and that in turn is a viable
possibility only if there is no moduli problem.
The two sets of papers therefore represent mutually exclusive
scenarios for the early Universe, and only time will tell which
if either is correct.

The rest of this paper is divided into two main sections plus a
concluding one. In Section 2 we study the effective potential
expected for flatons, both in the early Universe and in the
present era when it reduces to the ordinary low energy
effective potential. Special attention is paid to the case of moduli,
which is different from that of other flatons because the moduli potential
vanishes if supersymmetry is unbroken. The flaton decay rate is also
estimated. The reheat process for homogeneous flaton
oscillations is considered, taking account of possible parametric resonance.
In Section 3 a systematic account is given of the history of the
Universe, assuming that thermal inflation occurs and that there is a
moduli problem. The concluding section summarizes the results, and
points to future directions of research.

\section{Flat potentials and flatons}
\label{flat}

In a generic supersymmetric gauge theory there will be a large number of
directions in the space of the complex scalar fields\footnote
{Each scalar field is complex in supersymmetric theories because supersymmetry
relates it to the two degrees of freedom associated with a left- or
right-handed spin-half field. In this paper we are assuming that the fields
are canonically normalized in the regime of interest. (One cannot in general
canonically normalize the fields exactly over an extended region of field
space.)}
in which the potential $V$ is exactly flat, before supersymmetry
breaking and non-renormalizable terms are taken into account.
(This is true, for example, in the Minimal Supersymmetric Standard Model.)
After these effects are taken into account the potential is still almost
flat, in the sense that the energy scale $|V''|^{1/2}$
specifying the curvature of the potential is only of order
$10^2$ to $10^3\,{\rm GeV}$, out to field values many orders of magnitude
bigger than this scale.
In this paper we are interested in flaton fields, which by definition
correspond to flat directions with a nonzero vev.
The central theme of this paper is that flaton fields are cosmologically
significant, because they typically lead to thermal inflation, and because
they in any case oscillate homogeneously until a relatively late epoch.

A field with a nonzero vev is by definition
either a higgs field or a gauge singlet.
We will focus on the latter case in this paper,
since a straightforward interpretation of
the data indicate that the vev of the GUT higgs field is of order
$10^{16}\,{\rm GeV}$ which is too high to give viable thermal inflation.
Note, though, that in some GUT models there are additional higgs fields
with much smaller vevs \cite{Mohapatra}.

The cosmology of a given flaton field is largely determined by the form
of its effective potential. One needs to know both the low energy
effective potential which is relevant at the present era,
and the effective potential in the early Universe. Also, since the
case of moduli is somewhat different from that of flatons in general
we treat the moduli in a separate subsection after the general
discussion.

\subsection{The low energy effective potential}
\label{lep}

Consider a complex flaton field $\phi$.
In the limit where the potential is absolutely flat there is a
global $U(1)$ symmetry under
the transformation $\phi \to e^{i\alpha}\phi$, with an arbitrary choice
for the origin of $\phi$. In the full theory
this symmetry may survive
for one choice of the origin, at least to a good approximation,
 or it may be so badly broken as to be unrecognizable.

\subsubsection*{Global $U(1)$ symmetry}

We begin by considering the case where the symmetry survives.
Extensions of the Standard Model can indeed contain
spontaneously broken global $U(1)$ symmetries, a well known example being the
Peccei-Quinn symmetry associated with the axion
\cite{kimrev,Linde,KT,ouraxion,chunlukas}.
We initially suppose that the $U(1)$ symmetry is exact.
The potential then depends on $\phi$ only through $|\phi|$, and assuming
an effective theory that is valid right up to the Planck scale,
the potential in the flat direction is typically of the form
\begin{equation}
\label{V}
V = V_0 - m_0^2 |\phi|^2 +
\sum_{n=1}^\infty\lambda_nM_{\rm Pl}^{-2n}|\phi|^{2n+4}
\label{flatpot}
\end{equation}
The $|\phi|^2$ term
comes from soft supersymmetry breaking, which means that
$m_0\sim 10^2$ to $10^3\,{\rm GeV}$, and the higher order
terms are non-renormalizable terms.
The dimensionless couplings $ \lambda_n$
are at most of order 1, if the theory is indeed valid up to the
Planck scale.

The crucial feature of this potential, which distinguishes it from the
potential of a generic field and makes it flat, is the absence of
a term $\lambda |\phi|^4$ with $\lambda \sim 1$.
Such a term can be forbidden by discrete or continuous gauge symmetries,
in combination with supersymmetry.
Supersymmetry breaking then generates a $\lambda |\phi|^4$ term
with a suppressed coupling $\lambda \sim (m_0/M_{\rm Pl})^2$.
Such a term is negligible for flaton fields which are not moduli
and we have lost nothing by omitting it from Eq.~(\ref{flatpot}).
(The case of moduli will be discussed in a moment, and in more detail in
Section 2.5.)

As the notation suggests, we have in mind the case where
the mass-squared at the origin, $-m_0^2$,
is negative. This means that
the vev of $|\phi|$ does not vanish but rather
has a value $ M \gg m_0$.
To estimate $M$, suppose first that all of the
$\lambda$'s are of the same order. Then as one increases $|\phi|$,
the $|\phi|^6$ term comes in first, leading to
$ M = (3\lambda_1)^{-1/4} m_0^{1/2} M_{\rm Pl}^{1/2}
\sim \lambda_1^{-1/4} \times 10^{10}$ to $10^{11}\,{\rm GeV} $.
Now suppose instead that this term is negligible, so that
the $|\phi|^8$ term comes in first. Then
$ M = (4\lambda_2)^{-1/6} m_0^{1/3} M_{\rm Pl}^{2/3}
\sim {\lambda_2}^{-1/6} \times 10^{13}\,{\rm GeV} $.
If more terms are absent the vev will be raised further so the
predicted range is
$M\gtrsim 10^{10}\,{\rm GeV}$.
In the entire regime $ |\phi| \lesssim M $ the curvature $|V''|^{1/2}$ of the
potential is only of order $m_0$, which is of order $10^2$ to $10^3\,{\rm
GeV}$.
In particular the mass $m$ of the flaton particle is
of this order, and from now on we shall generally use it instead of $m_0
$ when writing down order of magnitude estimates.
The requirement $V(M)=0$ gives $V_0\sim m^2 M^2$, corresponding to
\begin{equation}
\left(\frac{V_0 ^{1/4}}
{10^6\,{\rm GeV}}\right) \sim \left(\frac{M}{10^{10}\,{\rm GeV}}\right)^{1/2}
\label{vzero}
\end{equation}
If the $n$th term dominates in Eq.~(\ref{flatpot}), then\footnote
{The mass-squared of the flaton particle is $\frac12 V''(M)$
because the canonically normalized complex field $|\phi|$ is related to
the canonically normalized real flaton particle field $\delta\phi$
by $|\phi|=M+\delta\phi/\sqrt2$.}
\begin{eqnarray}
m^2&=&2(n+1)m_0^2\\
M^{2n+2}M_{\rm Pl}^{-2n} &=& [2(n+1)(n+2)\lambda_n]^{-1} m^2
\label{vev}\\
V_0&=& [2(n+2)]^{-1} m^2 M^2
\end{eqnarray}

Rather than the non-renormalizable terms being suppressed by the Planck scale,
they might be generated by integrating out particles with GUT scale masses and
so instead be suppressed by $ M_{\rm GUT} \simeq 2 \times 10^{16}\,{\rm GeV} $.
This would correspond to taking $\lambda_n\lesssim
(M_{\rm Pl}/M_{\rm GUT})^{2n}$
and would give the somewhat looser lower bound
\begin{equation}
M\gtrsim 10^9\,{\rm GeV}
\end{equation}

We noted a moment ago that in Eq.~(\ref{flatpot}) the $|\phi|^4$ term
has a coupling
$\lambda\sim (m/M_{\rm Pl})^2$ which is many orders of magnitude less than
1. It may happen that the same
is true of one or more further terms. But for a flaton which is not
a modulus one expects to find, at not too high order,
a term whose coupling $\lambda_n$ is {\em not} many orders of magnitude
less than 1. As a result, one expects the vev of flaton which is not
a modulus to be several orders of magnitude below $M_{\rm Pl}$.
By contrast one expects for a
modulus that all couplings are strongly
suppressed, because the potential of a modulus vanishes
exactly when supersymmetry is unbroken. A natural order of magnitude for the
couplings of a modulus is $(m/M_{\rm Pl})^2$ making the vev of order
$M_{\rm Pl}$, though there are other possibilities. We shall discuss moduli in
more detail in Section 2.5.

The flat potential Eq.~(\ref{flatpot}) is not at all what cosmologists
generally
assume when they consider spontaneous symmetry-breaking in the early Universe.
Rather they assume, as for instance in the textbooks
\cite{Linde,KT,vs} and the reviews \cite{kimrev,hk}
that the potential is like the Standard Model higgs'
potential\footnote
{In the case of the Standard Model $\phi$ is a doublet and the symmetry is
$SU(2)$ but this is an irrelevant complication for our purpose.}
\begin{equation}
V = \lambda (|\phi|^2-M^2)^2
\label{vhiggs}
\end{equation}
with $\lambda\sim 1$.
For the Standard Model higgs, whose vev is of order $10^2\,{\rm GeV}$,
this potential is indeed natural from the viewpoint of
supergravity; it simply corresponds to a {\em non-flat\/} direction, in which
there is a $\lambda|\phi|^4$ term.
But when $M$ is much bigger than $m$ it becomes far less natural, and in our
view Eq.~(\ref{flatpot}) rather than Eq.~(\ref{vhiggs})
 should be regarded as the
`default' case.

So far we have taken the $U(1)$ symmetry to be exact, so that the
goldstone boson corresponding to the angular direction is massless.
If the symmetry is broken the goldstone boson will acquire a mass.
This mass is by definition much less than that of the flaton if the
symmetry is only slightly broken. On the other hand, as we now discuss
the symmetry may be strongly broken which means that the would-be
goldstone boson becomes just another flaton particle.

\subsubsection*{No $U(1)$ symmetry}

As a simple example, consider the superpotential $W=(\lambda/4M_{\rm
Pl})\phi^4$
with $\lambda\sim 1$.
After supersymmetry breaking the corresponding potential is of the form
\begin{eqnarray}
V(\phi)
& = & V_0 - m_0^2 |\phi|^2
+ \left( AW + B\phi \frac{\partial W}{\partial \phi}
	+ \mbox{c.c.} \right)
+ \left| \frac{\partial W}{\partial \phi} \right|^2 \\
\label{Vnosym}
& = & V_0 - m_0^2 |\phi|^2
+ \left( \frac{\lambda C\phi^4}{M_{\rm Pl}} + \mbox{c.c.} \right)
+ \frac{|\lambda|^2 |\phi|^6}{M_{\rm Pl}^2}
\end{eqnarray}
with $m_0$ and the magnitudes of
$A$, $B$ and $C$ all of order $10^2$ to $10^3\,{\rm GeV}$.

In this example $U(1)$ has been broken down to $Z_4$
(which leaves $\phi^4$ invariant), and there are four vacua
each with the same vev $|\phi|=M\sim |\lambda|^{-1/2}m_0^{1/2}M_{\rm
Pl}^{1/2}$.
In a given vacuum there are now two particles with mass
$10^2$ to $10^3\,{\rm GeV}$; one of them is the one corresponding to the radial
oscillation that we considered before, and the other is the would-be
goldstone boson corresponding to the angular oscillation.
We shall generally refer to them both as flatons.
Note that in the regime $|\phi|\ll M$ the $U(1)$ symmetry is approximately
restored, since the term $-m_0^2|\phi|^2$ dominates.

The $Z_4$ symmetry surviving in this example has ensured that there
are no linear terms in the expansion of $\phi$ about the origin,
and this feature
will become crucial when we consider the effective potential in
the early Universe. Of course any $Z_n$ symmetry will do for this
purpose, and it does not need to be exact.

In our discussion $-m_0^2$ has been taken to be negative.
If it is positive the potential has a minimum at the origin.
If this is also the position of the vev (ie., if it is the absolute
minimum) then the field is not a flaton and does not concern us.
It can however happen, as for instance in the model of \cite{false}, that the
origin corresponds to a false vacuum, with higher order terms generating
a large vev so that we are dealing with a flaton.
Thermal inflation with such a flaton is viable only if tunneling
to the true vacuum is rapid, which is typically not the case.

For simplicity we shall from now on
make frequent use of the notation appropriate to the
case where there is a $U(1)$, writing the potential as a function
only of $|\phi|$ and using $m$ to denote the mass of the flaton
particle.

\subsection{The flaton decay rate}
\label{decay}

There is a general expectation that
a flaton particle corresponding to oscillations around a vev $M$
will couple only weakly to particles with mass
much less than $M$. In particular, one expects
\cite{decay,yam,therm,Ross,yam2,interm,gutti}
that the flaton decay rate $\Gamma$
is at most of order $m^3/M^2$.

Consider first the decay into a pair of identical spin zero particles which
correspond to a real field $\psi$, with the renormalizable
effective interaction
$ \lambda |\phi|^2 \psi^2 $. Setting $|\phi|$ equal to its
vev this interaction gives a contribution $2\lambda M^2$ to the mass-squared
$m^2_\psi$.
Barring a precise cancellation,\footnote
{It has been pointed out to us by G. Dvali that such a
cancellation does occur in an SU(5) GUT where the doublet-triplet splitting
problem is `solved' by a fine tuned cancellation. In such a case
the decay rate
has the unsuppressed value $\Gamma\sim m$. When the problem is solved in a
more acceptable way this need not be so, but we will not pursue the
point here because our main focus is not on the GUT.}
it follows that
\begin{equation}
\lambda\lesssim \frac12\left(\frac{m_\psi}{M}\right)^2
\label{lambdabound}
\end{equation}
where the right hand side is at most $\frac18(m/M)^2$ or the decay would be
forbidden by energy conservation.
Substituting $|\phi|=M+\delta\phi/\sqrt2$, one finds that the flaton
decay rate corresponding to this interaction is
\begin{equation}
\Gamma=\frac{\lambda^2}{8\pi} \left(\frac{M}{m}\right)^2 m
\sqrt{1- 4 m_\psi^2/m^2 }
\label{decayrate}
\end{equation}
Maximizing this expression subject to the constraint
 Eq.~(\ref{lambdabound})
gives $ \Gamma  \lesssim 10^{-4} m^3/M^2$.

This effective interaction with a coupling of order
$\lambda\sim (m/M)^2$ is quite natural. For
instance an interaction $|\phi|^2 X^2$ might give some
field $X$ a mass of order $M$, and then an interaction
$\psi^2 X^2$ would generate it
through the diagram with a single $X$ loop.

For an effective interaction involving more powers of the fields and/or
derivatives the arguments are generally less precise, but one
expects suppression because such terms are non-renormalizable
and therefore involve inverse powers of some scale $\tilde M$ which is
presumably at least of order $M$. Consider for instance a term
involving one power of $\phi$ and two of $\psi$, with $2n$ derivatives.
Its coefficient is expected to be at most of order $\lambda' M^{-2n}$
with $\lambda'\sim M$, and
since the energy of all particles is of order $m$ (in the $\phi$
rest frame) this gives the decay rate Eq.~(\ref{decayrate})
with $\lambda\sim (m/M)^{2n}$.
For $n>1$ this is much smaller than
the upper limit Eq.~(\ref{lambdabound}), but for
$n=1$ it is bigger by a factor $(m/m_\psi)^2$ leading to
$\Gamma\sim (8\pi)^{-1} m^3/M^2$.
On the basis of this discussion, we shall assume that
\begin{equation}
\Gamma  = 10^{-2} \gamma m^3/M^2
\label{gamma}
\end{equation}
with $\gamma\lesssim 1$.

\subsubsection*{The decay into goldstone bosons}

A definite example of a derivative coupling is provided
by the decay of the `radial' flaton into the `angular' flaton,
or goldstone boson. Near the vev, the canonically normalized radial field
$s$ and angular field $a$ are defined by
\begin{equation}
\phi=(\frac s{\sqrt 2}+M) \exp(ia/\sqrt2 M)
\end{equation}
Expanding the canonical kinetic term ${\cal L}_{\rm kin}=
\partial_\mu\phi^* \partial^\mu\phi
$ to first order in $s$, one finds the canonical kinetic terms for
$s$ and $a$ plus an interaction term
\begin{equation}
{\cal L}_{\rm int}=\frac s{\sqrt 2M} \partial_\mu a \partial^\mu a
\end{equation}
The coefficient is of the advertised form $\lambda'/M^2$, with
$\lambda'=M/\sqrt 2$.

The goldstone bosons produced by this coupling can be cosmologically
dangerous, because their interaction can be too weak to thermalize
them. This will be discussed in connection with the axion in
\cite{david} (see also \cite{chunlukas}).

\subsubsection*{The flaton freeze-out temperature}

Though we have focussed on the decay rate, similar considerations apply
to collision rates. The rates for collisions involving a flaton and
other light particles are suppressed at energies well below $M$,
and therefore the freeze-out temperature below which flaton
particles cease to be in thermal equilibrium is very roughly
of order $M$. Note that this applies only in the true vacuum,
where the flaton field is oscillating about the vev.

\subsection{The effective potential in the early Universe}

In the early Universe, the interaction of a given field
with other fields will alter the
effective potential of that field, and in particular
the effective flaton potential $V(\phi)$ will be altered.

We should first clarify what is meant by the `effective potential
$V(\phi)$'. There is in reality a single effective potential
$V(\phi,\psi,\dots)$, which is a function of all the scalar fields.
It is natural to define the effective potential of any individual field
as the full potential with
all other fields held at their vevs, and this is the definition
that we had in mind for the low energy effective potential $V(\phi)$.
However in the early Universe all sufficiently light
scalar fields are significantly
displaced from their vevs, either homogeneously in the manner we have been
discussing for flatons, or inhomogeneously as for instance if the field
is in thermal equilibrium. Instead of evaluating the full effective potential
$V(\phi,\psi,\dots)$ with the other fields at their vevs one should
set them equal to their current time-averaged values, so that for
instance a term $\psi^2\phi^2$ is replaced by $\langle \psi^2 \rangle
\phi^2$. In addition, the actual form of the full effective potential
is affected by the presence of
particles with nonzero spin and also by kinetic terms, so that the
effective potential $V_{\rm early}(\phi,\psi,\dots)$ in the early Universe
is different from the low energy effective potential
$V_{\rm low}(\phi,\psi,\dots)$ which applies at present.
For both of these reasons, the
effective potential $V_{\rm early}(\phi)$ in the early Universe
is different from the low energy effective
potential $V_{\rm low}(\phi)$ which applies at present.

Although the form of the effective potential $V(\phi)$
changes with the history of
the Universe, its gradient will always vanish at the origin
provided that it is invariant under at least a $Z_n$ symmetry.
This tends to be at least approximately true in simple models, and
we shall take it for granted in what follows. Let us pause briefly
though to see why such a symmetry is common.
If the full potential $V(\phi,\psi,\dots)$
is expanded as a power series in all of
the fields each individual term will be invariant under one or more
$Z_n$ symmetries unless it consists of just the first power of one
field. For instance the term $\phi^2\psi^2$ is invariant under
a $Z_2$ acting on $\phi$, and another acting on $\psi$.
As we discussed in Section 2.1, only a few leading terms will be
important in practice, so it is reasonable that one or more
$Z_n$ symmetries will be approximately present in
the full potential.
Then the question of whether or not
the potential $V(\phi)$ of an individual field
possesses an approximate
$Z_n$ symmetry depends on the form of the full potential,
but again this is not unreasonable.

Taking it for granted that the gradient
of $V(\phi)$ vanishes at the origin, let us ask what is the
effective mass-squared $V''(0)$ in the early Universe.
(We continue to assume for simplicity that there is a
$U(1)$ symmetry, so that $V$ is a function only of $|\phi|$.)

First consider the era of ordinary inflation.
It has been known for some time \cite{dinefisch,coughlan,Ross,fvi}
that by looking at the form of the
full potential predicted by $N=1$ supergravity
one can identify contributions of order $\pm H^2$
to the mass-squared of every field. For the inflaton field(s)
these contributions have to cancel because otherwise inflation will not
occur, but for a generic field one does not expect a cancellation.
Assuming that flatons are not inflatons, the conclusion is that
their mass-squared during inflation is (at least) of order $\pm H^2$.

After inflation it is not so clear what the mass-squared will be.
In the extreme case where the interaction is of only gravitational
strength one expects a contribution of
the same order, $\pm H^2$ \cite{Dine}.
We noted earlier that in the true vacuum, the interaction of flaton
particles with other light particles is suppressed,
so one at first sight expects something like this estimate to
hold for a flaton field.
However, that suppression occurs because the vev of the flaton field
is large (the flaton particles correspond to small oscillations around
the vev). Near the origin the flaton field can have unsuppressed
interactions with light fields.

To see why, take as an example the interaction $\frac12\lambda|\phi|^2\psi^2$
that we considered earlier. When $\phi$ is at its vev this gives a
contribution $\lambda M^2$ to $m_\psi^2$.
Barring cancellations, $\lambda$ must therefore be small if $m_\psi$ is small.
But suppose that in contrast $m_\psi$ is of order $M$ and is
{\em generated} by this interaction. Then there is a
coupling $\lambda\sim 1$, and for flaton field values near the origin
the $\psi$ field becomes light. The result is that near the origin
the flaton field has an unsuppressed interaction with the light field $\psi$.

If $\phi$ is a higgs field, charged by definition
under a gauge symmetry, a coupling of this kind to at
least the gauge bosons and gauginos is inevitable. In the case where $\phi$
is neutral under all gauge symmetries, which is our focus here,
such a coupling is not inevitable but it is still quite natural;
for instance,  in models of the kind discussed in
\cite{flataxion,ewan,david}
a flaton field couples in this way to the right handed neutrino
and sneutrino.

Assuming that the flaton field near the origin
indeed has unsuppressed interactions with one or more particle species
having effective mass of order $|\phi|$, it will be in thermal equilibrium in
the regime $|\phi|\lesssim T$. (The upper limit comes from the fact
that at a given temperature particles with mass bigger than $T$
become too rare too maintain thermal equilibrium.)
The finite temperature
correction to the effective potential gives
the flaton an effective mass-squared \cite{decay,tomislav}
of order $(T^2-m_0^2)$, which gives
the effective potential a local minimum at the origin
for $T$ bigger than some critical temperature $T_{\rm C}\sim m_0\sim
m$. (As usual, $-m_0^2$ denotes the effective zero-temperature
mass-squared at the origin, and $m$ denotes the flaton particle
mass which is the parameter we normally focus on. Recall that both
$m_0$ and $m$ are of order $10^2$ to $10^3\,{\rm GeV}$.)

In addition to the local minimum at the origin, the effective potential
retains its true minimum at $\phi= M$ except at very high temperatures
$T\gtrsim M$, but there is no significant tunneling between the
two \cite{decay,tomislav}.

To summarize this discussion, if the flaton field has gravitational
strength interactions its mass-squared is expected to be of
order
$\pm H^2$. If, on the other hand, it has unsuppressed interactions
then it will be in thermal equilibrium in the
regime $|\phi|\lesssim T$ and in this regime there will be a necessarily
positive mass-squared of order $T^2\sim (M_{\rm Pl}/H) H^2$.
These are the most important possibilities for the effective
mass-squared but others exist, especially during inflation where one
might have a coupling to the inflaton field, say of the form
$\psi^2\phi_{\rm inf}^2$
(in particular, hybrid inflation \cite{hybrid} makes essential use of
such a coupling). As in this example, the positivity of the potential
tends to require that such a coupling again gives a positive
mass-squared.

\subsection{The cosmology of fields with flat potentials}

In the light of what we have done so far
there are the following four
possibilities for the cosmology of a field with a flat potential.

\paragraph*{(i) The field sits at the origin.}

If the minimum of the potential is at the origin
throughout the history of the Universe
then the field will sit there
apart from thermal and quantum
fluctuations. In that case it
does not undergo homogeneous oscillations
in the early Universe, and we are not concerned with it here.
It will in general have unsuppressed
interactions (at least if it is not a modulus)
and the corresponding particle species
will be produced through particle
collisions and decays involving these interactions.

\paragraph*{(ii) The field oscillates about the origin.}

Now suppose that although the minimum of the low energy effective potential
is at the origin, the minimum in the early Universe is displaced
because there is a
negative mass-squared of order $-H^2$.
In that case the field will start to
oscillate about the origin at the epoch $H\sim m$.
The oscillation is generally short lived, because the particles
corresponding to it generally have unsuppressed couplings
(except perhaps for moduli).
If there is no thermal inflation the oscillation
can however lead to viable baryogenesis
through the
Affleck-Dine mechanism \cite{DRTbaryo}.

\paragraph*{(iii) Thermal inflation occurs.}

In the two remaining cases the vev is nonzero,
so that we are dealing by definition with a flaton field.
Thermal inflation, which is the focus of the present paper,
occurs if the flaton field is held at zero in the
early Universe by the finite temperature.
It ends when the temperature falls to some critical
value $T_{\rm C}\sim m$ (provided
that the zero-temperature effective potential has no
barrier separating the origin from the vev),
after which the flaton field starts to oscillate about the vev.
The oscillation around the vev might persist for a long time because
the coupling of flaton particles to other light particles is
suppressed (Section 2.2, and Section 2.6 below).

\paragraph*{(iv) Flatons not leading to thermal inflation.}

In the fourth case the flaton field fails to be held at the origin
by the finite temperature of the early
Universe. This will occur if the flaton has an effective
mass-squared $\sim -H^2$ which
prevents it from ever being near the origin.
It will also occur whatever the sign of the mass-squared, if
the interaction of the flaton field is suppressed even
near the origin.
When $H$ falls to a value of order the flaton mass
$m$, the field starts to oscillate
about the vev, with an initial amplitude of order $M$.
(The initial amplitude is equal to $M$ if the initial field value is at
the origin. If the field is displaced from the origin by a mass-squared of
order $-H^2$, its value is typically of order $M$
when the field starts to
oscillate.) As in the previous case the oscillation might last for a
long time.

\subsection{The moduli potential}

What we have done so far, including the summary of the last
subsection, applies in essence to all flaton fields including
any which are moduli. On the other hand, moduli do have some properties
which distinguish them
from other scalar fields (`matter fields') and as a result the general
discussion acquires a somewhat different flavour when applied to them.

The low energy effective potential of a modulus vanishes exactly if
supersymmetry is unbroken. After supersymmetry breaking its potential
is generally thought to be flat, so that its curvature $|V''|^{1/2}$
is everywhere of order $10^2$ to $10^3\,{\rm GeV}$
(except near points of inflexion).
If a modulus has a nonzero vev, then as we discuss in a moment
its vev is generally expected to be of order $M_{\rm Pl}$.
To a large extent its properties can then be obtained simply
by setting $M=M_{\rm Pl}$ in formulas that apply to flatons in general,
but there are some special features.
These arise because one is forced to consider
field variations of order $M_{\rm Pl}$, in contrast with matter fields where
one need only consider much smaller variations (typically of order the
vev $M \ll M_{\rm Pl}$ for a flaton field which is not a modulus).

In order to talk about a nonzero vev for any field there has to be a
well defined origin, which will be defined as a point which is
invariant (`fixed') under the group of symmetries under which the field
transforms.
For matter fields this defines a unique origin, such that the symmetry
group consists of linear operators in field space.
For moduli the symmetries are more complicated, and there are in general
an infinite number of fixed points with a separation of order $M_{\rm Pl}$
(though only a finite number are physically distinct because the
symmetry is a discrete gauge symmetry).
If the vev of a modulus is at a fixed point it is natural to say that it
vanishes, and otherwise it is natural to define the vev as the distance
to the nearest fixed point.
These are the conventions that we have had in mind, without explicitly
stating them. The statement that the vev of some  modulus is of order
$M_{\rm Pl}$ just means that it is not close to any particular fixed point.
As with other fields, a modulus can have unsuppressed interactions
with other light fields only if it is close to a fixed point.

Each of the four possibilities for the cosmology of  a flaton field
listed in the last subsection exists for a modulus.
If possibility (i) holds for all moduli then there
is no moduli problem. Assuming that this is not the case, let us
look at the expected form of the effective potential
of a modulus $\Phi$.
For simplicity we will pretend that $\Phi$ is real, and take it to be
canonically normalized.
Before supersymmetry breaking is taken into account the potential
$V(\Phi)$ vanishes. With the breaking taken into account the
potential in the true vacuum (the low energy potential)
is generally thought to be of the form
\begin{equation}
V_{\rm true} = M_{\rm S}^4 \, f\left(\frac{\Phi}{M_{\rm Pl}}\right)
= \frac{1}{2} m_\Phi^2 (\Phi-\Phi_1)^2 + \ldots
\label{susyv}
\end{equation}
Here the supersymmetry breaking scale $M_{\rm S}$ is related to the
scale $m\sim 10^2$ to $10^3\,{\rm GeV}$ by
$ M_{\rm S} \sim (mM_{\rm Pl})^{1/2} \sim 10^{10}$ to $10^{11}\,{\rm GeV}$, and
$f(x)$ is a function whose value and low order derivatives are
typically of order
1 in the regime $|x|\lesssim 1$. We have expanded the potential
about its vev $\Phi_1$. Note that
the potential vanishes in the limit $M_{\rm S}
\to 0$ of unbroken supersymmetry, in accordance with the fact that we
are dealing with a modulus.

In the early Universe there will be additional supersymmetry breaking
because of the nonzero energy density $\rho$,
leading to an additional contribution to the potential of the form
\cite{dinefisch,coughlan,fvi,Dine}
\begin{equation}
V_{\rm cosm} = \rho \, g\left(\frac{\Phi}{M_{\rm Pl}}\right)
= \frac{\alpha}{2} H^2 \left( \Phi - \Phi_2\right)^2 + \ldots
\label{cosm}
\end{equation}
The function $g(x)$ has value and low order derivatives of order
1 (making $\alpha\sim 1$). The minimum of this potential is located at a
different value $\Phi_2$, which is displaced from the true vev
$\Phi_1$ by a distance $ \Phi_0 = \Phi_2-\Phi_1 \sim M_{\rm Pl}$.

We have in mind the case where
both $\Phi_1$ and $\Phi_2$ are nonzero
(case (iv) of the last subsection)
and of order $M_{\rm Pl}$ .
If $\Phi_2=0$ but $\Phi_1\neq 0$ (case (iii))
there is also a moduli problem, but it might be
rendered insoluble by domain walls (though in
analyzing this possibility within a given model one will have to remember that
the discrete symmetries under which the moduli transform are gauge
 symmetries). If
$\Phi_1=0$ but $\Phi_2\neq 0$ (case (ii))
there is no moduli problem if the relevant moduli
have unsuppressed couplings near the origin.

Although Eq.~(\ref{susyv}) is the simplest possibility for the
potential of a modulus there are others, which could lead to a vev
below the Planck scale. For example, if supersymmetry breaking
is due to hidden sector gaugino condensation
then the moduli potential might include terms of the form
$\mu^n |\phi|^{m+4} /M_{\rm Pl}^{n+m}$ where $\mu$ is related to
vevs arising from gaugino
condensation. These terms still vanish when
supersymmetry is unbroken, as is required for a modulus, but they
might generate a vev below the Planck scale.
For example the GUT Higgs could be a Wilson line modulus, with a vev
of order $10^{16}\,{\rm GeV}$ generated in this way \cite{flatguts}.
In considering the moduli problem we assume in this paper that at least some
moduli have a vev of order $M_{\rm Pl}$.

\subsection{The flaton reheat temperature}

Let us quantify the statement that the flaton field oscillations
in the early Universe last for a long time.

The oscillation of a flaton field with vev $M$
has initial amplitude $\phi_0\sim M$.
The corresponding energy density is
$\rho_\phi \sim \frac12 m^2 \phi_0^2$,
and the number density of the flaton particles is
$n_\phi \sim \frac12 m\phi_0^2$.
These particles have no random
motion because the field is homogeneous, so they constitute matter as
opposed to radiation. If the flaton is associated with thermal
inflation, the oscillation commences after thermal inflation and
immediately dominates the energy density. If not, the
oscillation
commences at the earlier epoch $H\sim m$, and
may or may not come to dominate the energy density.

If the
oscillation amplitude decreased like $a^{-3/2}$,
where $a$ is the scale
factor of the Universe, then the
energy per comoving volume of the flaton field would be conserved.
In fact, the energy drains away through the interactions of the flaton
field so that the oscillation amplitude decreases faster.

If the oscillation amplitude is sufficiently small and the interactions
are sufficiently weak, each flaton particle decays independently
so that the rate at which the energy drains away is simply
the particle decay rate $\Gamma $.
It has practically all disappeared soon after the time
\begin{equation}
\Gamma ^{-1}
\simeq 3 \gamma^{-1} 10^{-9} \left( \frac{M}{10^{11}\,{\rm GeV}} \right)^2
\left( \frac{300\,{\rm GeV}}{m} \right)^3 \,{\rm secs}
\end{equation}
where we have used Eq.~(\ref{gamma}).
Setting this time equal to $H^{-1}$ and
assuming that the decay products thermalize promptly we arrive at an
estimate of the `reheat temperature',
\begin{equation}
T_{\rm D}
\simeq g_\ast^{-\frac{1}{4}} \Gamma ^{\frac{1}{2}} M_{\rm Pl}^{\frac{1}{2}}
\simeq 3 \gamma^{\frac{1}{2}} \left( \frac{10^{11}\,{\rm GeV}}{M} \right)
\left( \frac{m}{300\,{\rm GeV}} \right)^{\frac{3}{2}} \mbox{GeV}
\label{TD}
\end{equation}
where $g_\ast \sim 10^2$ is the effective number of species at
$T=T_{\rm D}$.\footnote
{The following results will be used without comment in the text.
The entropy density of radiation at temperature $T$ is
$ s = (4/3)\rho/T=(2\pi^2/45) g_* T^3 = 1.01 g_*^{1/4} \rho^{3/4}$, where
$g_*(T)$ is the effective number of particle species in thermal
equilibrium, and $ \rho = (\pi^2/30) g_* T^4 $ is the energy density.
As the Universe expands the scale factor $a$ increases. The energy
density in relativistic particles (radiation) is proportional to
$a^{-4}$ and that in non-relativistic particles is proportional
to $a^{-3}$. In thermal equilibrium the entropy
$a^3 s$ in a comoving volume is constant and so is $g_*^{1/3}a T$.
According to the Standard Model, $g_*^{1/4}$ is in the range 1 to 2 for
$T \lesssim 100\,{\rm MeV}$, and then rises sharply to
become $\simeq 3$, finally
rising to $\simeq 4$ when $T\gtrsim 10^3\,{\rm GeV}$ in
supersymmetric extensions
of the Standard Model. We use the appropriate value in our estimates.}

As has been discussed recently in connection with ordinary inflation,
the assumption that each flaton particle decays independently
need not be correct \cite{brand,Kofman,Shtanov,Boy,yosh}
(see also \cite{ovrut,Ross}).
Instead, parametric resonance effects can drain away much of the
oscillation
energy as soon as the oscillation starts, leaving behind only some
fraction to decay at the single particle decay rate.
The energy drained away
goes initially into the creation of marginally relativistic
scalar particles. All species are produced which have sufficient
coupling to the flaton, including the flaton itself.
(We are not aware of any discussion of the possibility of the production
of bosons with spin 1 or higher through parametric resonance and it
may be that this also occurs. Fermions are not produced in
significant number because of Pauli blocking.)
If nothing happens to the produced scalar particles they will become
non-relativistic after a few Hubble times, and are expected to decay
at their one-particle decay rate.\footnote
{When the the particles have become non-relativistic one might
think that parametric resonance will recommence, since
the wavenumber of the corresponding scalar field is negligible compared
with its frequency. However, the collection of non-relativistic
particles corresponds to a {\em superposition} of almost-classical
quantum states, not to any one such state, since the phases of
the corresponding fields are uncorrelated, so it is not clear that the
parametric resonance formalism applies. More importantly, the amplitude
of the would-be classical oscillation
will typically be too small for parametric resonance to occur.
We are indebted to A.~D.~Linde for helpful correspondence about this
issue.} If, on the other hand, they thermalize then they turn into
highly relativistic radiation.

At the present time it is not clear whether parametric resonance can
really create particles which thermalize successfully.
However, it {\em is } clear that
the flaton component of the produced particles cannot thermalize
because here one knows that the interaction is too weak.
Furthermore, one expects that the energy density of the produced flatons
will be a significant fraction of the total energy density
\cite{andreipersonal}. Thus, even if the other produced particles
thermalize promptly one expects that a significant fraction of
non-thermalized energy will remain, and that a significant fraction
of {\em that} energy will be in flaton particles.

Any thermalized radiation produced by parametric resonance will redshift away,
so independently of the details one expects that a
few Hubble times after the end of thermal inflation
the energy density is dominated by non-relativistic
scalar particles, including the flatons and perhaps other species.
Each species will decay at the single-particle decay rate, so
we expect eventually to find only the longest-lived species, which
dominates the energy density until it decays.

For simplicity we shall assume in
what follows that this species is the flaton itself, and we
shall also ignore the effect of any radiation produced
by particle decay.
Thus we are in effect assuming that soon after thermal inflation
has ended, some fraction $\epsilon$ of the energy is in non-relativistic
flaton particles which decay according to the one-particle decay rate,
with the remainder in thermalized radiation. This should describe the
real situation at least approximately, provided that any non-flaton
particles produced decay at least as rapidly as the flatons.
The important special
case $\epsilon=1$ is considered, and the possibility that $\epsilon$ may
be very small is not discounted. However, as as discussed above,
this latter case seems unlikely because
one expects that parametric resonance
will convert a  significant fraction of the energy density into
flaton particles which interact too weakly to thermalize.

The upshot of this discussion is that despite the possible occurrence of
parametric resonance, one expects that the eventual reheat temperature
after thermal inflation is still the temperature $T_{\rm D}$ calculated from
the single-particle decay rate, as given by Eq.~(\ref{TD}).
If $T_{\rm D}$ is indeed the  reheat temperature, the requirement that it
be not too low places strong restrictions on $M$.
In order not to upset nucleosynthesis one must have
$T_{\rm D} \gtrsim 10\,{\rm MeV}$,
which requires $M \lesssim 10^{14}\,{\rm GeV}$
(taking $m<10^3\,{\rm GeV}$).\footnote
{In \cite{gutti} we estimated $M\lesssim 10^{16}\,{\rm GeV}$. The extra factor
100 came from three different sources. First we used the very naive
estimate $\Gamma\sim m^3/M^2$, corresponding to $\gamma^{1/2}=10$.
Second, we set $g_*^{1/2}=1$
where as the true value is more like $10^{1/2}$. Third, we rounded up
our estimate of $T_{\rm D}$ to the nearest power of ten which meant
multiplying it by of order $10^{1/2}$. It so happened that each
of these approximations went the same way to give the factor $100$.}
However, if $R$ parity is respected as is usually supposed,
there is a stable LSP which imposes a much stronger constraint.
Indeed, to bring
the LSP into thermal equilibrium so that it is not over-produced
(and can naturally have the correct abundance to be the dark
matter), one needs $T_{\rm D} $ substantially in excess of the
LSP decoupling temperature which is of order $1\,{\rm GeV}$.
Thus one needs $M\lesssim 10^{12}\,{\rm GeV}$.
Finally, one might wish to generate baryon number through the electroweak
transition which would require
$T_{\rm D} \gtrsim 100\,{\rm GeV}$ corresponding to
$M\lesssim 10^{10}\,{\rm GeV}$.
In view of the fact that these limits are perhaps rather conservative
(since one expects $\gamma$ to be significantly less than 1, and does
not anticipate $m$ as high as $10^3\,{\rm GeV}$) this last requirement
is hardly likely to be satisfied, but other baryogenesis mechanisms
exist as discussed in \cite{ewan}.

\section{Cosmology with thermal inflation}
\label{cwti}

We now give a systematic account of the history of the early Universe
in the case where there is thermal inflation.
We assume
that there is a moduli problem because this
provides the strongest motivation for thermal inflation, and assume
that at least some of the moduli have a vev of order $M_{\rm Pl}$.
We also assume that any radiation produced by parametric resonance
promptly thermalizes. With these assumptions there are the following eras
which we shall consider in turn.
\begin{enumerate}
\item Ordinary inflation.
\item Matter domination by the homogeneous oscillation of the inflaton
(unless full reheating occurs promptly).
\item Full reheating, which leads to radiation domination if it occurs
before the moduli start to oscillate.
\item Homogeneous oscillation of the moduli, starting
at the epoch $H\sim m_\Phi$. If reheating has previously occurred
there is now matter domination by the moduli. If it has not
occurred the moduli
and inflaton matter densities are roughly comparable, and remain so
until full reheating (of the inflaton matter). We assume
that full reheating takes place before the beginning of thermal inflation.
\item Thermal inflation.
\item Matter domination by the homogeneous oscillation of the
flaton field which caused thermal inflation
(unless reheating occurs promptly).
\item Full reheating of the flaton matter, leading to radiation domination
before nucleosynthesis after which the history of the Universe
is the standard one.
\end{enumerate}

\subsection{Before thermal inflation}

One expects the Universe to start with
an era of ordinary inflation \cite{Linde,KT},
whether or not there is a later epoch of thermal inflation.
During this era, the energy density
$\rho$ is dominated by the potential $V$ of the scalar fields, with
all except the inflaton field (or fields) fixed.
The inflaton field slowly rolls down
the potential, because in its direction the flatness conditions
$|M_{\rm Pl} V'/V|\ll 1$ and $|V''|\ll H^2$ are satisfied \cite{Linde,KT}.
We noted
earlier that in the context of supergravity
the second of these conditions
requires cancellations. Although these might be accidental
it is attractive to suppose that they
occur by virtue of some symmetry. One suitable symmetry
(most easily implemented in the context of hybrid inflation
\cite{hybrid}) was suggested in \cite{fvi,iss,mhi} and another
has been proposed in \cite{Murayama}.
A third possibility is to
invoke a global $U(1)$ symmetry as in \cite{natural},
but this is problematical because
the inflaton potential vanishes
in the limit where the symmetry is exact so that the magnitude
of $V''$ is difficult to control.\footnote
{An alternative idea \cite{rosssarkar} is to suppose that the
potential is exactly flat (or at least much flatter than that of the
inflaton field) in the direction of at least one field,
say a modulus, which couples to the inflaton.
The inflaton potential then depends on the value of this field,
which will vary from place to place in the Universe
allowing the possibility that we live in a region where the
inflaton potential happens to be sufficiently flat.
But this just pushes back to another level the problem of
finding cancellations which keep the potential flat in some direction.}

To avoid generating too much large scale cmb anisotropy
the potential at the end of ordinary inflation must
satisfy  \cite{bound}
\begin{equation}
V^{1/4} \lesssim 10^{16}\,{\rm GeV}
\label{vinfbound}
\end{equation}

At some epoch after ordinary inflation `reheating' occurs, which by
definition means that practically all of the energy density thermalizes
(except for the contribution of moduli).
If reheating is prompt the reheat temperature
is $T_{\rm R} \sim (V/g_\ast)^{1/4}$.
A naive estimate of the time taken for reheat
would be that it is the decay time
of a single inflaton particle, which typically
leads to a much lower reheat temperature.
However prompt conversion of a large
fraction of the energy density into marginally relativistic
particles is likely. In the commonly discussed case where
inflation ends with the oscillation of
a homogeneous inflaton field this is expected to occur through
the parametric resonance effect that we considered already for the case
of thermal inflation. It is also expected
to occur in the case of
hybrid inflation though a quantitative account of this case has not
yet been given, and will be more complicated because spatial
gradients are probably important from the beginning
\cite{fvi}. As we discussed earlier
these marginally relativistic particles may then
thermalize promptly leading to full or partial reheating.\footnote
{Note, though, that an extremely low fraction of the energy density
cannot thermalize
because thermalization requires that the interaction
rate per particle exceeds $H$.
If the decay products are charged under some gauge symmetry, this requires
$\alpha g_* T \gtrsim H$ where $\alpha $ is the gauge coupling.
Setting $\alpha\sim g_*^{-1/2}\sim 10^{-1}$, one finds
that it is satisfied only
if the fraction is bigger than
$V/(10^{16}\,{\rm GeV})^4$. This constraint does not seem to have been
noted before in the literature.}

Moduli
(more precisely, those moduli if any which are flatons with $M\sim
M_{\rm Pl}$)
are produced both before and after thermal inflation, and we shall call
the moduli from these sources respectively big bang moduli and thermal
inflation moduli.

When $H \gg m_\Phi$ the modulus' potential is given by
Eq.~(\ref{cosm}),
so that $\Phi$ is shifted from its true vacuum value by
$\Phi_0 = \Phi_2 - \Phi_1 \sim M_{\rm Pl}$.
$\Phi_2$ will depend on the composition of the Universe and so
$\Phi_0$ will change at any phase transitions, such as the end
of inflation, but $\Phi$ will rapidly settle down to its new minimum
as it is critically damped.
However, at the epoch $H\sim m_\Phi$ it starts to oscillate about the
minimum of its low energy effective potential, and after $H$ has fallen
significantly below $m_\Phi$, the oscillations will no longer be critically
damped and so are much more dangerous.

During thermal inflation $H\ll m_\Phi$, so the
effective potential is dominated by $V_{\rm true}$ but
$V_{\rm cosm}$ still gives a small contribution, so the position of the
minimum is shifted slightly from the true vacuum value.
Oversimplifying a bit, we
can estimate the shift by adding together $V_{\rm true}$
and $V_{\rm cosm}$ which gives
\begin{eqnarray}
V & = & \frac{1}{2} m_\Phi^2 \left( \Phi - \Phi_1 \right)^2
	+ \frac{\alpha}{2} H^2 \left( \Phi - \Phi_2 \right)^2
	+ \ldots \\
\label{vmod}
 & = & \frac{1}{2} m_\Phi^2 \delta\Phi^2
	+ \frac{\alpha}{2} H^2 \left( \delta\Phi - \Phi_0 \right)^2
	+ \ldots \\
\label{dmod}
 & = & \frac{1}{2} \left( m_\Phi^2 + \alpha H^2 \right)
\left( \delta\Phi - \frac{\alpha H^2}{m_\Phi^2 + \alpha H^2} \Phi_0
	\right)^2 + \ldots
\end{eqnarray}
where $ \delta\Phi = \Phi - \Phi_1 $ is the displacement of $\Phi$ from
its vev. In the last line $\alpha$ is of order 1, so the
minimum of the modulus' potential is shifted during thermal
inflation by an amount of order $ (H/m_\Phi)^2 M_{\rm Pl} $ \cite{gutti}.

To estimate roughly the abundance of big bang moduli,
we can assume that the modulus field starts to oscillate about its vev
when
$H\sim m_\Phi$ with amplitude of order $\Phi_0 \sim M_{\rm Pl}$.
The energy density $\rho_\Phi \sim m_\Phi^2 \Phi_0^2/2$
is of order the total energy density.
If reheating has already occurred
one can crudely set the radiation energy density equal to the total
energy density which leads to
the estimate
\begin{equation}
\label{bbm}
\frac{n_\Phi}{s} \sim \frac{\Phi_0^2}{10M_{\rm Pl}^{3/2}m_{\Phi}^{1/2}}
\end{equation}
(In this expression $s$ is the entropy density, and we are using the
standard results summarized in the footnote after Eq.~(\ref{TD}).)
If reheating occurs later
the moduli energy density is a fixed fraction of the total until
reheating,
and again setting the radiation density equal to the total
density after reheating one finds
\begin{equation}
\label{bbmh}
\frac{n_\Phi}{s} \sim \frac{\Phi_0^2 H_{\rm R}^{1/2}}
{10M_{\rm Pl}^{3/2}m_\Phi}
\end{equation}
It is described in the Appendix how a more sophisticated
calculation leads to the same results.

We shall assume that full reheating occurs before the onset
of thermal inflation (except for the contribution of moduli).
The opposite case will be discussed in \cite{ewan}.

These estimates for the moduli  apply to any flaton not giving
rise to thermal
inflation (option (iv) of Section 2.4), if $\Phi_0$ is replaced
by $M$.

\subsection{Thermal Inflation}

Thermal inflation will occur if one or more of the
flaton fields is trapped at
the origin in the early Universe.
For the moment we suppose that
only one is trapped.

The trapping may initially be due to a non-thermal contribution to the
mass-squared such as that of order $H^2$.
However if full reheating occurs before the beginning of thermal
inflation
then well within a Hubble time of the end of inflation enough entropy to
trap the flaton at zero will have been released even by the single
particle
decay of the inflaton.

If full reheating is indeed delayed to the epoch when
thermal inflation begins, the temperature at that epoch is
of order $g_*^{-1/4}V_0^{1/4}\sim (mM)^{1/2}$ corresponding to
\begin{equation}
\left(\frac{T}{10^6\,{\rm GeV}}\right)\sim
\left(\frac{M}{10^{10}\,{\rm GeV}}\right)^{1/2}
\end{equation}
At the other extreme where reheating occurs before the moduli start to
oscillate, the temperature at the beginning of thermal inflation
is reduced by a factor $(M/M_{\rm Pl})^{1/6}$.
During thermal inflation
$T\propto \exp(-Ht)$ and it ends at $T=T_{\rm C}\sim m$, so there
are at most of order $\frac12 \ln(M/m)\sim 10$ e-folds of thermal inflation.
This will not much
affect the cosmological density perturbation generated about
50 e-folds before the beginning of ordinary inflation,
though there might be a slight change in
the spectral index.

\subsection{Entropy production after thermal inflation}

After thermal inflation ends, relic radiation from the first
hot big bang plays no further role.
The flaton field now starts to oscillate around its vev with
initial amplitude $M$, corresponding to non-relativistic
flatons (matter) which dominate the energy density.

The decay of the flaton field generates entropy.
If there is no parametric resonance the entropy per comoving volume
increases linearly from the end of thermal inflation until the
flaton decays, leading to an increase in the entropy by a factor
\begin{equation}
\Delta \sim
\frac{ 4 V_0 / 3 T_{\rm D} }{ (2\pi^2/45) g_\ast(T_{\rm C}) T_{\rm C}^3 }
\sim \frac{ V_0 }{ 75 T_{\rm D} T_{\rm C}^3 }
\label{delta1}
\end{equation}

Now suppose instead that there is parametric resonance which promptly
thermalizes a substantial fraction of the energy density,
leaving
a fraction $\epsilon$ in the flatons.
This will increase the entropy by a factor
\begin{eqnarray}
\label{dpr}
\Delta_{\rm PR}
& \sim & \frac{ g_\ast(T_{\rm PR})^{1/4} (1-\epsilon)^{3/4} V_0^{3/4} }
{ (2\pi^2/45) g_\ast(T_{\rm C}) T_{\rm C}^3 }
\sim \frac{ (1-\epsilon)^{3/4} V_0^{3/4} }{ 25 T_{\rm C}^3 } \\
& \sim & 10^{10}
\left( \frac{ M }{ 10^{10}\,{\rm GeV} } \right)^{\frac{3}{2}}
\left( \frac{ m }{ T_{\rm C} } \right)^3
\left( \frac{ 300\,{\rm GeV} }{ m } \right)^{\frac{3}{2}}
\left( \frac{ V_0 }{ m^2 M^2 } \right)^{\frac{3}{4}}
\end{eqnarray}
The radiation energy density may initially
dominate, but we assume that it falls below that of the residual
flatons before the epoch $T_{\rm D}$.
The entropy release from the decay of these flatons
is significant only during the era
$ T_{\rm D}\lesssim T\lesssim ( T_{\rm D}^4 T_{\rm eq})^{1/5} $ \cite{KT},
so it is a good approximation to regard
this  entropy release as suddenly occurring at the epoch
$T\sim T_{\rm D}$.
It increases the entropy by a further factor
\begin{eqnarray}
\label{dd}
\Delta_{\rm D} & \sim & \frac{ 4 \epsilon V_0 / 3 T_{\rm D} }
{ g_\ast(T_{\rm PR})^{1/4} (1-\epsilon)^{3/4} V_0^{3/4} }
\sim \frac{ \epsilon V_0^{1/4} }{ 3(1-\epsilon)^{3/4} T_{\rm D} } \\
& \sim & 10^{6} \epsilon
\left( \frac{ M }{ 10^{10}\,{\rm GeV} } \right)^{\frac{1}{2}}
\left( \frac{ \mbox{GeV} }{ T_{\rm D} } \right)
\left( \frac{ m }{ 300\,{\rm GeV} } \right)^{\frac{1}{2}}
\left( \frac{ V_0 }{ m^2 M^2 } \right)^{\frac{1}{4}}
\end{eqnarray}
The total entropy increase is
\begin{eqnarray}
\label{d}
\Delta
& \sim & \Delta_{\rm PR} \Delta_{\rm D}
\sim \frac{ \epsilon V_0 }{ 75 T_{\rm D} T_{\rm C}^3 } \\
& \sim & 10^{15.5} \epsilon
\left( \frac{ M }{ 10^{10}\,{\rm GeV} } \right)^2
\left( \frac{ \mbox{GeV} }{ T_{\rm D} } \right)
\left( \frac{ m }{ T_{\rm C} } \right)^3
\left( \frac{ 300\,{\rm GeV} }{ m } \right)
\left( \frac{ V_0 }{ m^2 M^2 } \right)
\label{delta}
\end{eqnarray}

Eq.~(\ref{dd}) is only supposed to apply if it gives a value
$\Delta_{\rm D}$ bigger than 1, which fails to be true in
the small $\epsilon$ regime $\epsilon\lesssim T_{\rm D}/V_0^{1/4}$.
This is the regime in which the flaton oscillation fails to dominate the
energy density before it disappears at the epoch $T=T_{\rm D}$.
In it $\Delta_{\rm D}$ is practically equal to 1,
and $\Delta=\Delta_{\rm PR}$ has the $\epsilon$-independent value
given by Eq.~(\ref{dpr}).

We shall not consider the case where parametric resonance creates
radiation which fails to thermalize, and hence quickly reverts
to matter in the form of homogeneously oscillating scalar fields.

\subsection{Solving the moduli problem with single thermal inflation}

In order not to upset nucleosynthesis, the moduli abundance
$ n_\Phi / s $ must be less than $10^{-12} $ to $10^{-15}$
when nucleosynthesis begins \cite{constraint}. Let us see what is
required to satisfy this bound, first for the
big bang moduli  and then for the moduli
produced after thermal inflation.

We can assume that the flaton oscillation comes to dominate
the energy density, because the assumption
can be shown to be
valid in the regime of parameter space satisfying the nucleosynthesis
bound on the moduli abundance and to lead to an overestimate of the moduli
abundance outside this regime. As a result we can use Eq.~(\ref{d}),
and combining it with Eq.~(\ref{bbmh}) one finds
that the abundance of big bang moduli after
thermal inflation is
\begin{eqnarray}
\frac{n_\Phi}{s}
& \sim & \frac{ \Phi_0^2 H_{\rm R}^{1/2} }{ 10 m_\Phi \Delta M_{\rm Pl}^{3/2} }
\sim \frac{ 8 \Phi_0^2 H_{\rm R}^{1/2} T_{\rm D} T_{\rm C}^3}
{ \epsilon m_\Phi V_0 M_{\rm Pl}^{3/2} } \\
& \sim & 10^{-16}
\left( \frac{ 10^{12}\,{\rm GeV} }{ M } \right)^{\frac{3}{2}}
\left( \frac{ 1 }{ \epsilon } \right)
\left( \frac{ M_{\rm Pl} H_{\rm R} }{m_\Phi M} \right)^{\frac{1}{2}}
\left( \frac{ T_{\rm D} }{ \mbox{GeV} } \right) \nonumber \\
&& \times
\left( \frac{ T_{\rm C} }{ m_\Phi } \right)^3
\left( \frac{ \Phi_0 }{ M_{\rm Pl} } \right)^2
\left( \frac{ m_\Phi }{ 300\,{\rm GeV} } \right)^{\frac{1}{2}}
\left( \frac{ m_\Phi^2 M^2 }{ V_0 } \right)
\label{32}
\end{eqnarray}
In these formulas $H_{\rm R}$ is to be considered as being in the range
$ m_\Phi (M/M_{\rm Pl}) \lesssim H_{\rm R} \lesssim m_\Phi $.
The lower limit comes from our assumption that full reheating after
ordinary inflation occurs before the beginning of thermal inflation,
and if $H_{\rm R}$ actually exceeds the upper limit
the above formulas give the correct result when it is set equal to this
limit.

To analyze these constraints, assume first that $T_{\rm D}
\gtrsim 1\,{\rm GeV}$ as is required if the LSP is stable, and recall
that this implies $M\lesssim 10^{12}\,{\rm GeV}$, from Eq.~(\ref{TD}).
In Eq.~(\ref{32}), the round brackets in the second line are all of
order unity, so we see that
the big bang moduli may be sufficiently diluted for $M$
as low as $ 10^{9}\,{\rm GeV}$, though this requires all parameters to be
pushed to the limit and a more reasonable estimate of the lower limit
might be
$ 10^{11}\,{\rm GeV}$.
Now assume only that $T_{\rm D}\gtrsim
10\,{\rm MeV}$, as required by nucleosynthesis, which implies $M\lesssim
10^{14}\,{\rm GeV}$. Then we see that unless $\epsilon$ is very small
it should be possible to solve the moduli problem, with no significant
additional constraint on $M$.

Now consider the moduli produced after thermal inflation.
{}From Eq.~(\ref{dmod}), the minimum of the
potential during thermal inflation is displaced from its true vacuum minimum
by an amount $ \delta\Phi \sim ( V_0 / m_\Phi^2 M_{\rm Pl}^2 ) \Phi_0 $.
The dynamics at the end of thermal inflation will be complicated but one would
expect to generate a moduli number density
$ n_\Phi \sim m_\Phi \,\delta\Phi^2 / 2
\sim \Phi_0^2 V_0^2 / 2 m_\Phi^3 M_{\rm Pl}^4 $
at the end of thermal inflation.
Therefore the abundance of thermal inflation moduli is
expected to be
\begin{eqnarray}
\frac{n_\Phi}{s}
& \sim & \frac{ \Phi_0^2 V_0^2 / 2 m_\Phi^3 M_{\rm Pl}^4 }
{ g_\ast(T_{\rm PR})^{1/4} (1-\epsilon)^{3/4} V_0^{3/4} \Delta_{\rm D} }
\sim \frac{ \Phi_0^2 V_0 T_{\rm D} }{ 3 \epsilon m_\Phi^3 M_{\rm Pl}^4 } \\
& \sim & 10^{-15.5}
\left( \frac{ M }{ 10^{12}\,{\rm GeV} } \right)^2
\left( \frac{ 1 }{ \epsilon } \right)
\left( \frac{ T_{\rm D} }{ \mbox{GeV} } \right) \nonumber\\
&& \times \left( \frac{ \Phi_0 }{ M_{\rm Pl} } \right)^2
\left( \frac{ 300\,{\rm GeV} }{ m_\Phi } \right)
\left( \frac{ V_0 }{ m_\Phi^2 M^2 } \right)
\label{35}
\end{eqnarray}
Bearing in mind the relation between $T_{\rm D}$ and $M$, we see that
the abundance of thermal moduli does not impose a significant additional
constraint.

Moduli will be produced in the flaton's decay with abundance
\begin{equation}
\frac{n_\Phi}{s} \sim
\frac{ \Gamma_{\phi\rightarrow\Phi} }{\Gamma} \frac{n_\phi}{s}
\end{equation}
Since the flaton energy density is $mn_\phi$ and we are assuming that it
all thermalizes,
$n_\phi/s$ is of order $T_{\rm D}/m$ and therefore
\begin{equation}
\frac{n_\Phi}{s}
\sim 10^{-16} \left( \frac{ \mbox{GeV} }{ T_{\rm D} } \right)
\left( \frac{ \Gamma_{\phi\rightarrow\Phi} }{ m^3 / 8\pi
M_{\rm Pl}^2 } \right)
\left( \frac{m}{ 300\,{\rm GeV} } \right)^2
\end{equation}
which is probably sufficiently small.

Finally we consider the possible thermal creation of
gravitinos, moduli and modulinos after thermal inflation.
Gravitinos, for which the most detailed calculations exist,
appear to be created in a cosmologically safe abundance provided that
the maximum temperature is less than \cite{Leigh}
$ 10^9\,{\rm GeV} $, and a similar result presumably holds for moduli
and
modulini since in all cases the interaction with other particles is of
gravitational strength. This bound is satisfied after thermal inflation
even in the extreme case where
most of the energy density thermalizes immediately.

\subsection{Double Thermal Inflation}
\label{double}

So far we assumed that only one flaton field gives thermal inflation,
or in other words that only one flaton field has a thermal mass-squared
which traps it at the origin in the early Universe. If two
or more flaton fields are trapped the situation is in general much more
complicated, but it simplifies considerably if the fields do not
interact significantly. We treat this simple situation now,
leaving the case of interacting fields to future publications
\cite{david,ewan}.
Thus we consider
two flaton fields
$\phi_1$ and $\phi_2$, and assume that each of their potentials is of the form
Eq.~(\ref{V}),
\begin{equation}
V(\phi_1,\phi_2)= V_1+ V_2- m_1^2 |\phi_1|^2
- m_2^2 |\phi_2|^2
\end{equation}
The higher order terms stabilize the fields at $ \phi_i = M_i $,
and the constants $V_1$ and $V_2$ are the values of the separate
potentials at the origin, with $V_i \sim m_i^2 M_i^2$.
The critical temperatures at which the fields roll away from zero
are $T_{{\rm C}i}$, and we take $ T_{\rm C1} > T_{\rm C2} $.
When the temperature drops below $T_{\rm C1}$,
$\phi_1$ will roll away from zero.

If parametric resonance does not produce significant thermalization, the
second field now also rolls away promptly and the situation is not
substantially different from the case of thermal inflation. If on the
other hand a significant fraction of the energy density is thermalized
by parametric resonance, the temperature will be raised sufficiently
to trap the second field before it has a chance to roll away, leading to
a second epoch of thermal inflation driven by the potential
\begin{equation}
V = V_2 - m_2^2 |\phi_2|^2 + \ldots
\end{equation}

The residual flatons left after parametric resonance from the first
epoch of thermal inflation may be troublesome if they do not decay before
nucleosynthesis.
Their abundance evaluated after the second epoch of thermal inflation is
\begin{eqnarray}
\frac{n_{\phi_1}}{s}
& \sim & \frac{ \epsilon_1 V_1 / m_{\phi_1} }
{ g_\ast(T_{\rm PR1})^{1/4} V_1^{3/4} \Delta_2 }
\sim \frac{ 20 \epsilon_1 V_1^{1/4} T_{\rm D2} T_{\rm C2}^3 }
{ \epsilon_2 m_{\phi_1} V_2 } \\
& \sim & 10^{-14.5}
\left( \frac{ M_1 }{ 10^{14}\,{\rm GeV} } \right)^{\frac{1}{2}}
\left( \frac{ 10^{12}\,{\rm GeV} }{ M_2 } \right)^2
\left( \frac{ \epsilon_1 }{ \epsilon_2 } \right)
\left( \frac{ T_{\rm D2} }{ \mbox{GeV} } \right)
\nonumber \\
&& \times
\left( \frac{ T_{\rm C2} }{ m_{\phi_1} } \right)^3
\left( \frac{ m_{\phi_1} }{ 300\,{\rm GeV} } \right)^{\frac{1}{2}}
\left( \frac{ V_1 }{ m_{\phi_1}^2 M_1^2 } \right)^{\frac{1}{4}}
\left( \frac{ m_{\phi_1}^2 M_2^2 }{ V_2 } \right)
\end{eqnarray}
Thus, a second epoch of thermal inflation may significantly dilute
the residual flatons from a first epoch, which could remove the
restriction $M_1\lesssim 10^{14}\,{\rm GeV}$ which is otherwise demanded
by nucleosynthesis. Conceivably one may in this way make thermal
inflation viable with a GUT Higgs field or even with a modulus,
though more investigation is needed to see whether this is a real
possibility.

Henceforth we will assume that $M_1$ is sufficiently small to allow $\phi_1$
to decay before nucleosynthesis, which
allows us to take $M_2$ small enough to have a comfortably high final
reheat temperature.

\subsection{Solving the moduli problem with double thermal inflation}

For simplicity we take the abundance of big bang moduli before thermal
inflation to be $ n_\Phi / s \sim 0.1 \Phi_0^2 M_{\rm Pl}^{-3/2}
m_{\Phi}^{-1/2} $
and assume that parametric resonance leads to effectively complete reheating
so that $ \epsilon \simeq 0 $.
These assumptions lead to the maximum possible moduli abundance.
Each epoch of thermal inflation then increases the entropy by a factor
$ \Delta_i \sim \Delta_{{\rm PR}i} \sim V_i^{3/4} / 25 T_{{\rm C}i}^3 $.
Therefore the abundance of big bang moduli after double thermal inflation is
\begin{eqnarray}
\frac{n_\Phi}{s} & \sim &
\frac{ \Phi_0^2 }{ 10 m_{\Phi}^{1/2} \Delta_1 \Delta_2 M_{\rm Pl}^{3/2} }
\sim \frac{ 60 \Phi_0^2 T_{\rm C1}^3 T_{\rm C2}^3 }
{ m_{\Phi}^{1/2} V_1^{3/4} V_2^{3/4} M_{\rm Pl}^{3/2} } \\
 & \sim & 10^{-17.5}
\left( \frac{ 10^{13}\,{\rm GeV} }{ M_1 } \right)^{\frac{3}{2}}
\left( \frac{ 10^{10}\,{\rm GeV} }{ M_2 } \right)^{\frac{3}{2}}
\left( \frac{ T_{\rm C1} }{ m_\Phi} \right)^3
\left( \frac{ T_{\rm C2} }{ m_\Phi} \right)^3 \nonumber \\
&& \times
\left( \frac{\Phi_0}{M_{\rm Pl}} \right)^2
\left( \frac{ m_\Phi }{ 300\,{\rm GeV} } \right)^{\frac{5}{2}}
\left( \frac{ m_\Phi^2 M_1^2 }{ V_1 } \right)^{\frac{3}{4}}
\left( \frac{ m_\Phi^2 M_2^2 }{ V_2 } \right)^{\frac{3}{4}}
\end{eqnarray}

The abundance of thermal inflation moduli produced at the end of the first
epoch of thermal inflation evaluated after the second epoch of thermal
inflation is
\begin{eqnarray}
\frac{n_\Phi}{s}
& \sim & \frac{ \Phi_0^2 V_1^2 / 2 m_\Phi^3 M_{\rm Pl}^4 }
{ g_\ast(T_{\rm PR1})^{1/4} V_1^{3/4} \Delta_2 }
\sim \frac{ 3 \Phi_0^2 V_1^{5/4} T_2^3 }{ m_\Phi^3 V_2^{3/4} M_{\rm Pl}^4 } \\
& \sim & 10^{-16.5}
\left( \frac{ M_1 }{ 10^{13}\,{\rm GeV} } \right)^{\frac{5}{2}}
\left( \frac{ 10^{10}\,{\rm GeV} }{ M_2 } \right)^{\frac{3}{2}}
\left( \frac{ T_2 }{ m_\Phi } \right)^3 \nonumber \\
&& \times
\left( \frac{\Phi_0}{M_{\rm Pl}} \right)^2
\left( \frac{ m_\Phi }{ 300\,{\rm GeV} } \right)
\left( \frac{ V_1 }{ m_\Phi^2 M_1^2 } \right)^{\frac{5}{4}}
\left( \frac{ m_\Phi^2 M_2^2 }{ V_2 } \right)^{\frac{3}{4}}
\end{eqnarray}

The abundance of thermal inflation moduli produced at the end of the second
epoch of thermal inflation is
\begin{eqnarray}
\frac{n_\Phi}{s}
& \sim & \frac{ \Phi_0^2 V_2^2 / 2 m_\Phi^3 M_{\rm Pl}^4 }
{ g_\ast(T_{\rm PR2})^{1/4} V_2^{3/4} }
\sim \frac{ \Phi_0^2 V_2^{5/4} }{ 8 m_\Phi^3 M_{\rm Pl}^4 } \\
& \sim & 10^{-14}
\left( \frac{ M_2 }{ 10^{10}\,{\rm GeV} } \right)^{\frac{5}{2}}
\left( \frac{\Phi_0}{M_{\rm Pl}} \right)^2
\left( \frac{ 300\,{\rm GeV} }{ m_\Phi } \right)^{\frac{1}{2}}
\left( \frac{ V_2 }{ m_\Phi^2 M_2^2 } \right)^{\frac{5}{4}}
\end{eqnarray}

We see that two independent
bouts of thermal inflation can solve the moduli problem
for a wide range of the vevs, even if parametric resonance is extremely
efficient.

\subsection{Topological defects}

We end this paper with a brief discussion of the cosmological
production of topological defects,
namely walls, strings, monopoles and textures.

Each type of topological defect is associated with a scalar field
(in general multi-component) with nonzero vev. Among several
possibilities, we consider here only two cases. The first is that
the vev belongs to a GUT higgs potential, and that it has the
non-flat form usually considered. The second is that the vev belongs to
a flat potential.

For a GUT higgs with the standard non-flat potential
the temperature after inflation is never high enough for
the defects to form by the usual Kibble mechanism \cite{vs,hk}.
(We are not of course concerned with any defects forming before ordinary
inflation since they have been diluted away.)
They can only form near or at the end of
ordinary inflation, and even that requires that the
bound Eq.~(\ref{vinfbound}) on the inflationary potential
is saturated \cite{fvi}.

Consider first monopoles, using the standard results \cite{vs}.
The abundance of monopoles,
after some initial annihilation, settles down soon after the GUT
transition to a value
$n/s\sim 10^{-10}$. The strongest bound on their
present abundance comes
from baryon decay catalysis in neutron stars, which requires
$n/s\lesssim 10^{-37}$. Thus the entropy must increase by a factor
$10^{27}$ between the end of ordinary inflation and the present.
If reheating after ordinary inflation is
prompt, the factor is the one $\Delta$ arising from thermal inflation.
We see from Eq.~(\ref{delta}) that a single bout of thermal inflation is
probably insufficient, but two bouts could be enough. Alternatively, if
reheating after ordinary inflation is long delayed this gives an
additional increase
$\Delta_{\rm ord}\sim 10^{16}\,{\rm GeV}/T_{\rm R}$,
which could be enough to make
just one bout of thermal inflation viable.

Depending on the GUT symmetry, gauge strings might also be
produced, which would be cosmologically significant
perhaps providing candidates for the origin of large scale
structure. On cosmological scales their
evolution is not affected by thermal inflation because
their spacing is outside the horizon during that epoch.
(This is just the statement that there are much less than
50 $e$-folds of thermal inflation.) The same applies to
other defect networks formed before thermal inflation
(global domain walls, monopoles, strings
or textures).

Consider now defects associated with a flat potential.
They form if at all at the end of thermal inflation.
Consider first the case of
$Z_n$ symmetry (Sections 2.1 and 2.3).
A discrete symmetry
used to be regarded as problematical for cosmology,
because when it is spontaneously
broken it seems to lead to cosmologically forbidden
domain walls. However, if the
symmetry is also explicitly broken, as will typically be the case for
the flaton potential, there need be no problem
because walls do not necessarily form and if they do form they do not
necessarily survive (because the vacua on either side of a wall
may have different energy density). If, on the other hand, it is exact
it will probably
be a discrete gauge symmetry which again avoids the domain wall
problem because there is only one physical vacuum.

If there is a global $U(1)$ symmetry, strings can
form at the end of thermal inflation with the strings later joined by
walls if the symmetry is approximate.
An example of this might be
Peccei-Quinn symmetry \cite{david}. Local strings forming at the end
of thermal inflation would have
too little energy to be cosmologically significant.
Finally, if the flaton field giving rise to thermal inflation has
two or more components as in Section 3.5 then
monopoles or textures might form at the end of thermal inflation but we
have not considered this case.

\section{Summary and Conclusion}
\label{sum}

Flatons are scalar fields with masses $m$ of order $10^2$
to $10^3\,{\rm GeV}$ and
vacuum expectation values $ M \gg m$.
They arise naturally in supersymmetric
theories and indeed it is not unreasonable to suppose
that they are the only source of vevs in this range
(in the observable sector).
Flatons with $ M \gtrsim 10^{14}\,{\rm GeV} $ are
cosmologically dangerous, and in
particular moduli with $ M \sim M_{\rm Pl} $
are overproduced by twenty orders of
magnitude in the standard cosmology, which
is the well-known Polonyi/moduli problem.
In this paper we have explained how the problem may be solved by
flatons
with smaller vevs, in the range
$10^9\,{\rm GeV} \lesssim M \lesssim 10^{13}\,{\rm GeV}$ that
is theoretically very natural
for flatons other than moduli.

Such flatons solve the moduli problem by generating an era of thermal
inflation.
Thermal inflation occurs when the flaton is held at zero by thermal effects,
and it typically lasts for about 10 $e$-folds and occurs at a very low
energy scale.
These properties are precisely what is required to sufficiently
dilute the moduli produced before thermal inflation without affecting the
density perturbation produced during ordinary inflation (10 $e$-folds),
while not regenerating them again afterwards
(low energy scale).
Detailed calculations show that
a single epoch of thermal inflation driven
by a flaton
whose vev is within one or two orders of magnitude of $10^{12}\,{\rm GeV}$
can solve the moduli problem,
though the constraints are quite tight.

It is easier for thermal inflation to rescue flatons with vev's
$M \gtrsim 10^{14}\,{\rm GeV}$ (in particular, moduli with $M\sim M_{\rm Pl}$)
if the latter do not themselves give rise to thermal inflation.
Remarkably, segregation of flatons into a class which thermally
inflate and have small vevs, and a class which do not and have large
vevs is exactly what one expects from a theoretical viewpoint.
The larger the vev of a flaton, the less likely it is to be
trapped at the origin in the early Universe, because the finite
temperature contribution to the effective potential becomes ineffective
at field values bigger than the temperature.

There are several aspects of cosmology which we have not addressed in
the present paper, notably axion cosmology and baryogenesis
which will be the subjects of respectively \cite{david,ewan}.
Let us close by briefly discussing the latter topic.
As successful thermal inflation sufficiently dilutes any pre-existing
moduli it will also dilute any pre-existing
baryon number to negligible amounts.
However, as will be discussed in \cite{ewan},
there are several possibilities for baryogenesis within the context
of thermal inflation itself.
One especially promising mechanism can occur if
the flaton which gives rise to thermal inflation
also generates the mass of a right-handed neutrino.
A lepton asymmetry can then be generated after thermal inflation.
The partial reheat temperature after thermal inflation can be high
enough to restore the electroweak symmetry,
and so this lepton asymmetry can be converted into a baryon
asymmetry by non-perturbative electroweak effects \cite{leptogenesis}.

\section*{Appendix}

To arrive at a more sophisticated estimate of the moduli abundance
we solve the equation
of motion of the modulus in the potential Eq.~(\ref{vmod}), which is
\begin{equation}
\label{modeom}
\ddot{\delta\Phi} + 3 H \dot{\delta\Phi} + m_\Phi^2 \delta\Phi
+ \alpha H^2 \left( \delta\Phi - \Phi_0 \right) = 0
\end{equation}
where we take $H=p/t$ with $p=1/2$ for radiation domination and $p=2/3$ for
matter domination.
One would expect $\alpha \sim 1$.
An estimate of $\Phi_0$ can be obtained by taking the distance between
the self-dual points of the target space modular symmetry SL(2,{\bf Z})
\cite{Cvetic} using the usual orbifold K\"{a}hler metric for the moduli.
This gives $\Phi_0^2 \sim 0.1$.
In this case one can easily check that our approximation of neglecting the
contribution of the moduli to the energy density of the Universe before the
asymptotic solution is attained is consistent.
$\delta\Phi$ will rapidly settle to $ \delta\Phi = \Phi_0$
when $ H \gg m_\Phi $,
and so we take $ \delta\Phi(0) = \Phi_0$ and $\dot{\delta\Phi}(0)=0$.
With these initial conditions, Eq.~(\ref{modeom}) has the solution
\begin{equation}
\delta\Phi = \alpha p^2 \Phi_0
\left( \frac{1}{m_\Phi t} \right)^{\frac{3p-1}{2}}
s_{\mu,\nu}\left( m_\Phi t \right)
\end{equation}
where $s_{\mu,\nu}$ is a Lommel function, $ \mu = -3(1-p)/2 $ and
$ \nu^2 = - \alpha p^2 + (3p-1)^2 /4 $.
At late times
\begin{equation}
\delta\Phi \sim \frac{\alpha \Phi_0}{\sqrt{\pi}}
\left( \frac{p}{2} \right)^{\frac{4-3p}{2}}
\Gamma \left( \frac{1+\mu}{2} + \frac{\nu}{2} \right)
\Gamma \left( \frac{1+\mu}{2} - \frac{\nu}{2} \right)
\left( \frac{H}{m_\Phi} \right)^{\frac{3p}{2}}
\sin \left( m_\Phi t + \frac{(2-3p)\pi}{4} \right)
\end{equation}
The coefficient has a weak dependence on $\alpha$ and $p$, and for
$ \alpha \sim 1 $ and $ p = 1/2 $ or $2/3$ we get to a good approximation
\begin{equation}
\delta\Phi \sim \frac{4}{3} \Phi_0
\left( \frac{H}{m_\Phi} \right)^{\frac{3p}{2}}
\sin \left( m_\Phi t + \frac{(2-3p)\pi}{4} \right)
\end{equation}
Therefore the moduli abundance is given by
\begin{equation}
\frac{n_\Phi}{s} \sim \frac{\Phi_0^2}{10m_{\Phi}^{1/2}}
\left( \frac{H}{m_\Phi} \right)^{\frac{3(2p-1)}{2}}
\end{equation}
Setting $p=1/2$ gives Eq.~(\ref{bbm}) and setting $p=2/3$ gives
Eq.~(\ref{bbmh}).

\subsection*{Acknowledgements}
While this work was in progress EDS was supported by a
Royal Society Fellowship at Lancaster University.
EDS is now supported by the JSPS.
DHL acknowledges funding from PPARC and the EC,
and both authors acknowledge funding from
the Aspen Physics Institute
where the work was completed.
We are indebted to
Andrei Linde for extensive correspondence about parametric resonance,
and have benefited from useful discussions with
Graham Ross, Beatriz de Carlos, Andre' Lukas, Raghavan Rangarajan,
Lisa Randall and Steven Thomas.

\end{document}